\documentclass[aps,float,prd,psfig,amsmath,twocolumn]{revtex4}
\usepackage[dvipdfmx]{graphicx}
\usepackage{amsmath,amssymb,natbib,siunitx,booktabs,url}
\usepackage{color}

\newcommand{\lsim}{\lesssim}
\newcommand{\gsim}{\gtrsim}

\newcommand{\lmk}{\left(}
\newcommand{\rmk}{\right)}
\newcommand{\lnk}{\left\{ }
\newcommand{\rnk}{\right\} }
\newcommand{\lkk}{\left[}
\newcommand{\rkk}{\right]}
\newcommand{\lla}{\left\langle}
\newcommand{\p}{\partial}
\newcommand{\rra}{\right\rangle}

\newcommand{\beq}{\begin{equation}}
\newcommand{\beqa}{\begin{eqnarray}}
		  \newcommand{\eeq}{\end{equation}}
\newcommand{\eeqa}{\end{eqnarray}}

\newcommand{\bp}{{\bar{p}}}

\begin{document}

\title{Measuring the dipole component of
possible Galaxy-binary alignment in the mHz band}

\author{Naoki Seto }
\affiliation{Department of Physics, Kyoto University, 
Kyoto 606-8502, Japan
}

\date{\today}

\begin{abstract}

We discuss the usability of the gravitational wave detector LISA for  studying the orientational distribution of compact white dwarf binaries in the Galactic bulge.  We pay special attention to measuring the dipole pattern of the distribution around the Galactic rotation axis.  Based on  our new formulation, which  leverages the parity properties of the involved systems, 
we found that  the apparent  thickness of the bulge in  the sky becomes critical for the dipole measurement.  We also discuss the extra-Galactic studies for  black hole binaries and neutron star binaries with BBO/DECIGO.  

\end{abstract}


\maketitle

\section{introduction}
Over the past 100 years, many researchers have been interested in whether the orbital orientations of nearby binaries are correlated with Galactic structures (see e.g. \cite{chang} for early works). 
The astrometric data of  visual binaries have been the primary resources in this context.   We can find the recent observational situations on  nearby binaries  in the detailed study by Agati et al. \cite{agati}. Analyzing 95 binaries within 18pc from the Sun, they argued  that these binaries appear to be randomly oriented. 

Meanwhile,  in 2023, Tan et al. \cite{tan}  published a paper on the orientations of the symmetric axes of planetary nebulae (PNe) around the Galactic bulge. They selected 14 PNe that host (or are inferred to host) short-period ($\lsim 1$ day) binaries. They found that the symmetric axes of these PNe tend to be parallel to the Galactic plane. Since PNe are gases ejected during the formation of white dwarfs \cite{kwi}, the symmetric axes of the selected PNe are considered to align with the orbital axes of the inner short-period binaries. Therefore, the observed patterns of the PNe suggest that the associated 14 short-period binaries around the bulge are not randomly oriented \cite{tan}, in contrast to the aforementioned binaries around the Sun \cite{agati}. Binaries are fundamental systems in astrophysics \cite{Duchene:2013cba},  and  their formation process could be highly diverse even in the Galaxy.  We are encouraged to make further observational studies on orientations of binaries in various environments.

For the  orbital angular momentum orientation (denoted below by the unit vector $\vec j$) of  a Galactic  binary, considering the whole-part relation, one would be particularly interested in its possible statistical alignment with the Galactic rotation vector $\vec q$. More specifically, if we take the rotation vector $\vec q$ as the polar direction for the spherical harmonic expansion, the dipole pattern $(l,m)=(1,0)$ of the unit vector $\vec j$ would be a high-priority target. Unfortunately, the symmetric axes of the PNe are insensitive to the odd $l$ patterns, since the observed PN axes do not allow us to distinguish between $\vec j$ and $-\vec j$.  We will later comment on this point again, in relation to the parity properties of the spherical harmonics. Note that the two vectors $\vec j$ and $\vec q$ are both axial, and the harmonic expansion does  not depend on the  adopted handedness. 

The proposed space gravitational wave (GW) interferometer  LISA is expected to detect $\sim 10^4$ close white dwarf binaries (CWDBs) in the Galaxy \cite{LISA:2022yao,Hils:1990vc,Nelemans:2000es,Ruiter:2007xx,Breivik:2019lmt,Nissanke:2012eh,Korol:2017qcx,Lamberts:2019nyk} and determine their geometrical configuration \cite{Cutler:1997ta,Takahashi:2002ky}.   A significant fraction (e.g. $\sim 30\%$) of the CWDBs are estimated to be the bulge component \cite{Ruiter:2007xx}. 
It should also be stressed that, for LISA, we can easily control the selection effects, e.g. by using CWDBs only above $\sim 4$mHz, where LISA will make a complete Galactic survey \cite{LISA:2022yao,Seto:2022iuf} (see also \cite{gaia} for the selection effects of  the Gaia binary survey).  Therefore, as pointed out by the author  \cite{Seto:2024odc}, 
LISA could be  an essentially new observatory to examine the orientational distribution of Galactic binaries relative to the Galactic structure.   However, in the previous paper \cite{Seto:2024odc}, the author applied a distant observer approximation, in which the angular thickness of the bulge was ignored. In fact, as shown later,  this approximation  results in a blindness to the odd $l$-patterns, including the dipole mode. 
In this work, as a follow-on study, we pay a special attention to  the dipole pattern measurement, now carefully taking into account the  small  bulge thickness.  In addition, we newly introduce  appropriate estimators for simultaneously measuring various harmonic patterns. Our approach can be expanded to    extra-Galactic studies for binaries composed by  black holes and neutron stars.
We will  make a brief discussion on the  prospect  with BBO and DECIGO.

{This paper is organized as follows. }
In Sec. II, we discuss the axisymmetric model for the angular momentum distribution of the bulge CWDBs and make its spherical harmonic decomposition. In Sec. III, we explain the fourfold degeneracy at estimating the  polarization angle for a CWDB with LISA.  We apply a systematic scheme for handling the degeneracy. In Sec. IV, we introduce simple estimators to separately measure the observable spherical harmonic patterns of the orientational distribution function. We also present  the shot noise limits for the dipole and quadrupole measurements. {In Sec. V, we estimate the fluctuations induced by the parameter  estimation errors.  }
In Sec. VI, we discuss 
the extra-Galactic studies with BBO and DECIGO. Sec. VII is devoted to a short summary.  

\section{Distribution function}
 
 For CWDBs in the Galactic bulge, the joint probability distribution function for the orbital orientations $\vec j$ and the sky directions $\vec n$ can be generally expressed as ${\cal P}(\vec j,\vec n)$. Here, in view of the mixing processes discussed below,  we assume that they are independently distributed  ${\cal P}(\vec j,\vec n)={\cal P}(\vec j){\cal R}(\vec n)$ and the function ${\cal P}(\vec j)$ is axisymmetric around the Galactic rotation axis $\vec q$. 
 
 Given the probable bar-like structure in the central part of our Galaxy \cite{ga,ga2}, one might conversely expect that there should be strong azimuthal (in particular $m=\pm 2$) patterns in the function ${\cal P}({\vec j})$.  However, the formation epochs of the bulge CWDBs were likely to be more broadly distributed than the rotation  period of the bar  (currently $\sim 10^8$ yr \cite{ga,ga2}) \cite{Nelemans:2000es}.  
 Therefore, even if the orientations of the binaries have some azimuthal patterns ($m\ne 0$) at their formation epochs,  the chronological phase mixing should  largely reduce  the azimuthal  patterns in the integrated population at present.  In contrast,  the axisymmetric patterns $m=0$ would be more resistant to the mixing.
One could also expect positional intermixture facilitating independence between  $\vec j$ and $\vec n$, since the majority of bulge stars have not been corotating with the bar and also have random velocities \cite{ga,ga2,gd}.
In the present explanation, for simplicity, we supposed that the orientation vectors $\vec j$ had been  frozen  since the binary formation,  as discussed in \cite{tan}. 
Below, we mainly use the Galactic latitude $b$ and longitude $\eta$ to represent the binary position $\vec n$ in the sky, and put ${\cal R}({\vec n})={\cal R}(b,\eta)$.

The axisymmetric function ${\cal P}(\vec j)$ should depend only on the angle $\gamma$ defined by the product
$\cos \gamma={\vec j}\cdot {\vec q}$,  thus  containing only $m=0$ patterns  
\beq
{\cal P}({\vec j})=\sum_{l=0}^\infty a_{l0}Y_{l0}(\gamma)=\sum_{l=0}^\infty\sqrt{\frac{2l+1}{4\pi}} a_{l0}  L_l({\vec j}\cdot {\vec q}) \equiv {\cal P}_A(\cos\gamma) \label{pa}
\eeq
with the normalization  condition 
$a_{00}=1/\sqrt{4\pi}$.  The spherical  harmonics  $Y_{lm}$ are generally expressed as 
\beq
Y_{lm}(\theta,\phi)=\sqrt{\frac{2l+1}{4\pi}\frac{(l-m)!}{(l+m)!}} L_{lm}(\cos\theta)e^{im\phi}\label{aa6}
\eeq
with the associated Legendre polynomials $L_{lm}$, including the special cases $L_l\equiv L_{l0}$  for $m=0$  \cite{Jackson:1998nia}.  We present concrete examples for the Legendre polynomials $L_l$ 
 \beq
L_0=1,~ L_1(\alpha)=\alpha,~L_2(\alpha)=\frac{(3\alpha^2-1)}2.
 \eeq   
 The functions $L_{lm}$ satisfy the symmetric relations 
\beq
L_{lm}(-\alpha)=(-1)^{l+m}L_{lm}(\alpha),\label{aa10}~~
L_{l-m}(\alpha)=(-1)^m L_{lm}(\alpha) 
\eeq
 with the  identity $L_{lm}(1)=\delta_{m0}$. From Eq. (\ref{aa10}), we can easily show the parity relation 
 \beq
 Y_{lm}(-\vec j)=(-1)^l Y_{lm}(\vec j).\label{parity}
 \eeq

\begin{figure}
 \includegraphics[width=1\linewidth]{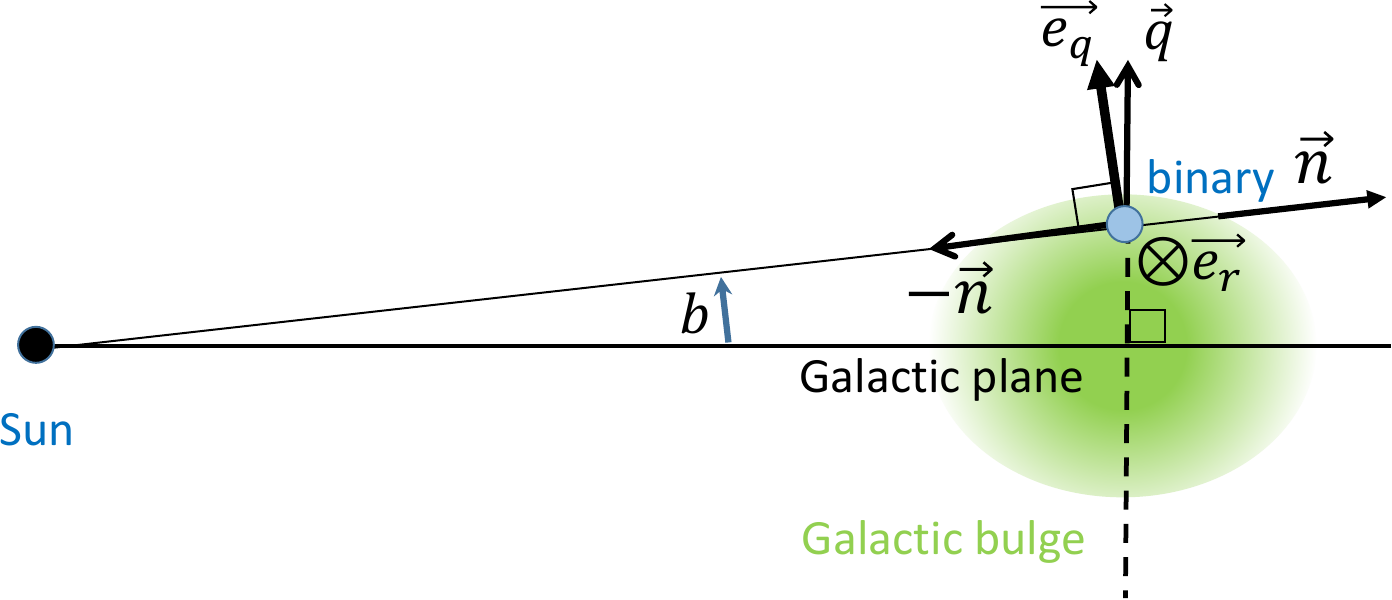} 
 \caption{Geometrical configuration of a binary (at the Galactic latitude $b$) on the slice of a constant Galactic longitude.  [\underline{Directional vectors}]: All four vectors (${\vec q}, {\vec n}, \vec e_q,\vec e_r)$ are unit vectors.
 The vector $\vec n$ shows the direction of the binary, and $\vec e_q$ is the transverse projection of the Galactic rotation axis  $\vec q$ with the offset angle $b$.  We also set $\vec e_r=\vec n\times \vec e_q$.
 [\underline{Source frame}]:  
  We define the source frame as the polar coordinate system with the polar direction $-\vec n$ (GW propagation direction) and set the reference direction  $\vec e_q$ for the
 azimuthal angle.  In this source frame, the orientation of the binary $\vec j$ (not shown here) is specified by the inclination (polar) angle $I$ and the polarization (azimuthal) angle $\psi$. 
 }  \label{fig:del}
\end{figure}

To discuss GW strain observation (particularly for Sec. III),  it is crucially advantageous to work in a polar coordinate system, where polar axis is directed in the GW propagation direction ($-\vec n$ in Fig. 1). In Fig. 1, there is still freedom to set the reference direction for the azimuthal angle of the polar coordinate and 
use the  projected vector $\vec e_q$, respecting the available Galactic symmetry.  We hereafter refer to this polar coordinate system as the source frame, and the orientation vector $\vec j$ will be decomposed into the polar angle $I$ and the azimuthal  angle $\psi$ (respectively called the inclination and polarization angles).  In the source frame, the Galactic axis $\vec q$ is given by  $(\pi/2+b,0)$ with the Galactic latitude $ b$. 

We next evaluate the orientational distribution  functions 
$p(I,\psi|b)$ for the two variables $(I,\psi)$ at a given sky direction $(b,\eta)$ (actually independent of the Galactic longitude $\eta$ due to the symmetry around $\vec q$).  We just need to rewrite the function ${\cal P}_A(\cos\gamma)$ in Eq. (\ref{pa}) in terms of the variables $(I,\psi)$, taking into account the offset angle $b$. Here the addition theorem for the spherical harmonics works quite efficiently. For two unit vectors $\vec x$ and $\vec y$, we generally have  \cite{Jackson:1998nia}
\beq
L_l({\vec x}\cdot {\vec y})=\frac{4\pi}{2l+1}\sum_{m=-l}^lY_{lm}(\vec x)Y^*_{lm}(\vec y). \label{add}
\eeq
Applying this theorem to each term $L_l({\vec j}\cdot \vec q)$ in Eq. (\ref{pa}) with the  angular parameters in the source frame, we obtain 
 \beqa
p(I,\psi|b)&=&\sum_{l=0}^\infty\sum_{m=-l}^la_{l0}\sqrt{\frac{4\pi}{2l+1}}Y_{lm}(I,\psi)Y^*_{lm}(b+\frac\pi2,0)\nonumber\\
&=&\sum^\infty_{l=0}\sum_{m=-l}^la_{l0}D_{lm}(b)Y_{lm}(I,\psi)
\label{pdf132}.
\eeqa
Here the real coefficients $D_{lm}(b)$ are defined by
\beq
D_{lm}(b)\equiv (-1)^{l+m} \sqrt{\frac{(l-m)!}{(l+m)!}} L_{lm}(\sin b) \label{dlm0}
\eeq
and
satisfy    $D_{l-m}=(-1)^mD_{lm} $.  From Eq. (\ref{aa10}), we can readily obtain the parity relations
\beq
D_{lm}(-b)=(-1)^{l+m} D_{lm}(b).
\eeq
 From Eqs. (\ref{pa}) and (\ref{pdf13}), the coefficients can  also be given as the inner products 
\beq
D_{lm}(b)= \int_{4\pi}dId\psi \sin I ~ Y_{l0}(\gamma)Y_{lm}^*(I,\psi),  \label{crs}
\eeq
which will be useful to geometrically understand the meaning of  the coupling coefficients $D_{lm}(b)$.  For each degree $l$,  we can   confirm  the power conservation  
\beq
\sum_{m=-l}^lD_{lm}(b)^2=1
\eeq
 (e.g. directly putting $\vec x=\vec y=\vec q$ in Eq. (\ref{add}) and using $L_l(1)=1$).

We present the explicit form for the dipole coefficient 
\beq
D_{10}(b)=-\sin b.
\eeq
In Fig. 2 (panels (a) and (b)), we provide its intuitive explanation.   For $|b|\ll 1$ (valid for our bulge CWDBs), we have $D_{10}(b)=O(b)$ and the most of the power is stored in the $m=\pm1$ modes with 
\beq
|D_{11}(b)|=|D_{1-1}(b)|\simeq 1/\sqrt2.
\eeq
  In general, we have $D_{lm}(b)=O(b)$ for odd $l+m$, and $D_{lm}(b)=O(1)$ for even $l+m$.

\begin{figure}[t]
 \includegraphics[width=0.9\linewidth]{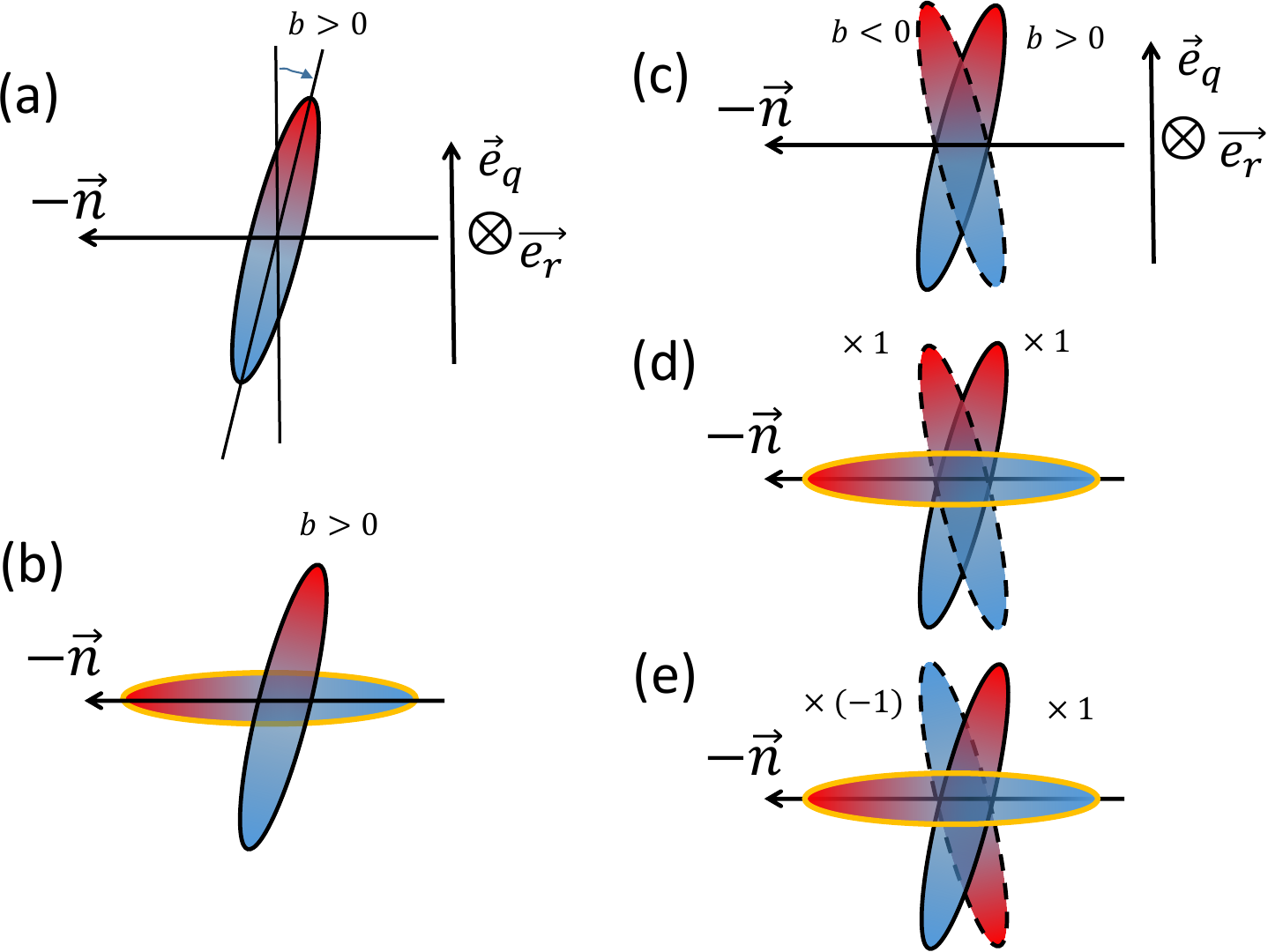} 
 \caption{Dipole patterns $Y_{10}(\gamma)$ in the source frame.  The red and blue respectively show positive and negative values. (a) For binaries at a  Galactic latitude $b$, the intrinsic dipole pattern $Y_{10}(\gamma)$ is tilted by the angle $b$.  (b) As shown in Eq. (\ref{crs}),  the coefficient $D_{10}(b)=-\sin b$ is  the cross product between the tilted profile $Y_{10}(\gamma)$ and the probe $Y_{10}^*(I)=Y_{10}(I)$ (edged with the yellow line),  which is axisymmetric around the polar direction $-\vec n$.  From the symmetries, we can easily see   $D_{10}(-b)=-D_{10}(b)$ with $D_{10}(0)=0$. (c) Dipole patterns from the pairwise latitudinal directions $\pm b$. (d) When combined in the same phase,  the total profile becomes antisymmetric in the directions $\pm \vec e_q$, and the product with the axisymmetric one $Y_{10}(I)$ vanishes with  $g_{100}=0$.  (e) When combined in the opposite phase, the total profile becomes vertically  symmetric,  resulting in $g_{101}\ne 0$.
 }  \label{fig:del}
\end{figure}

\section{Geometrical cancellations}

LISA will be a drastically new tool to geometrically explore the Galactic CWDBs.  However, observationally,  we cannot uniquely specify the polarization angle $\psi$ for  each bulge CWDB from its GW signal.  There are actually  four possible solutions (see e.g. \cite{Cornish:2003vj}).  Following the previous paper \cite{Seto:2024odc}, we explain the basic aspects of the fourfold degeneracy and apply a simple prescription for our pattern analysis.   Note that most of the CWDBs are expected to have negligible eccentricities, e.g. due to tidal dissipation effects.

First, let us consider a circular binary with the orientation angles $(I,\psi)$ and rotate it around the polar axis $-\vec n$ by $180^\circ$.  The new binary has the angles $(I,\psi+\pi)$, but its gravitational waveform is identical to that of the original one, because of the spin-2 nature of the wave \cite{pw}. Correspondingly, using only  GW data, we cannot determine which of the two values $\psi$ and $\psi+\pi$ is the true polarization angle.

In fact, for the CWDBs,  there will be  more spurious solutions other than $\psi+\pi$.  If we rotate the binary now by $90^\circ$ (namely $\psi\to \psi+\pi/2$), the polarization tensors change signs. For the lowest quadrupole signal, these signs can be absorbed as the overall phase shift, resulting in the signal degeneracy.  We can, in principle,  distinguish the two states $(I,\psi)$ and $(I,\psi+\pi/2)$ by using the post Newtonian effects (see e.g. \cite{pw,LIGOScientific:2020stg}).  Unfortunately, these relativistic effects are very weak for our bulge CWDBs in the mHz band and will be totally masked by the measurement noise of LISA.  Taking into account the spin-2 character again, there will be   fourfold degeneracy between $\psi,\psi+\pi/2,\psi+\pi$ and $\psi+3\pi/2$ (namely three spurious solutions). 

LISA will newly provide us with  orientational information for thousands of bulge  CWDBs.   At some stage, we need to appropriately address the fourfold degeneracy not just twofold. 
Here, as proposed in \cite{Seto:2024odc}, we identify the four solutions by  compactifying the range of the polarization angle $\psi$ from  the original one $[0,2\pi]$ to the folded one  $[0,\pi/2]$.
More concretely, we consider the following folded function 
\beqa
{\bar p}(I,\psi|b)&\equiv&  p(I,\psi|b) +p(I,\psi+\pi/2|b)\nonumber\\
& &+  p(I,\psi+\pi|b) + p(I,\psi+3\pi/2|b) \label{fold}
\eeqa
as our observable distribution. This is a systematic approach without ambiguities. Here it is essential to work in  the source frame. 

Importantly, the folding operation has a strong side effect.  In Eq. (\ref{fold}),  each harmonic term $Y_{lm}$  from Eq. (\ref{pdf132}) is now proportional to $\sum_{k=0}^3 \exp(im k\pi/2)$, which vanishes unless the order $m$ is multiple of 4.   Therefore, we can write down 
\beq
{\bar p}(I,\psi|b)=4\sum_{l=0}\sum_{m\in K}^{|m|\le l}a_{l0}D_{lm}(b)Y_{lm}(I,\psi)\label{pdf13}
\eeq
with the group 
\beq
K\equiv \lnk 0,\pm4,\pm8,\cdots\rnk.
\eeq
Indeed, the functions $2Y_{lm}$ ($m\in K$) compose an orthonormal basis for the folded orientational space $0\le I\le \pi$ and $0\le \psi\le \pi/2$.

As mentioned earlier, in  original function (\ref{pdf132}),  our target signal $a_{10}$ is mainly contained in the patterns $Y_{1,\pm1}$ with $|D_{1,\pm1}|\simeq 1/\sqrt2$ for the bulge directions at $|b|\ll 1$.  However, these strong patterns at $m=\pm1\notin K$  disappear at the folding operation. To detect the target $a_{10}$, we thus need to statistically amplify the weak available  pattern $\propto Y_{10}(I,\psi)$ with the coupling coefficient  $|D_{10}|=|\sin b|\ll 1$, allowed by the literal bulge structure in the sky.

To discuss the statistical amplification of the weak signals from bulge CWDBs at various Galactic  latitudes $b$, we define the joint PDF
\beq
\bp(I,\psi,b)=\bp(I,\psi|b) R(b). \label{mul}
\eeq
Here we put  the reduced distribution
\beq
R(b)\equiv \int_0^{2\pi} d\eta {\cal R}(b,\eta).
\eeq

 In the direction vertical to the Galactic plane, the density profile of bulge stars is known to be well approximated by a Gaussian distribution with the scale height $\sim 0.5$kpc (nearly independent of the Galactic longitude $\eta$) \cite{ga}. Given the distance to the Galactic center  $\sim 8.3$kpc,   we have the corresponding angular scale  $\sim \sigma\equiv 0.5/8.3=0.06$ and use the following model hereafter
\beq
R(b)=\frac1{\sqrt{ 2\pi\sigma^2}}\exp\lkk  \frac{-b^2}{2\sigma^2} \rkk . \label{gauss}
\eeq

At this point,  we will briefly comment on the geometrical studies on the planetary nebulae.  As mentioned earlier, their symmetric axes do not distinguish $\vec j$ and $-\vec j$.    
Then, given the parity relations in Eq. (\ref{parity}), the combination ${\cal P}(\vec j)+{\cal P}(-\vec j)$ is  insensitive to the odd $l$ patterns, including the dipole modes $l=1$.

\section{Anisotropy measurement}
Next, we outline the estimation of the anisotropic signals $a_{l0}$ from the observed data 
$\lnk I_i,\psi_i,b_i\rnk$ with the labels   $i=1,\cdots,N$ ($N$: the number of relevant binaries) and $0\le \psi_i\le \pi/2$.  

In \cite{Seto:2024odc}, the author discussed a primitive dualistic approach, targeting only the quadrupole pattern $a_{20}$. 
Here, for simultaneously measuring the multiple coefficients  $a_{l0}$, we introduce  more elaborate estimators (see also \cite{pro})
\beq
{\hat X}_{lmn}=\frac{1}N\sum_{i=1}^N b_i^n Y^*_{lm}(I_i,\psi_{i}). \label{estb}
\eeq
From Eqs. (\ref{pdf13}) and (\ref{mul}), their expectation values become
\beqa
\lla{\hat X}_{lmn} \rra_{\rm sample} &=&\lla b^n Y_{lm}^*(I,\psi) \rra_{I\psi b}\\
&=&a_{l0}g_{lmn}
\eeqa
with the coefficients
\beqa
g_{lmn}\equiv  \int_{-\pi/2}^{\pi/2} db R(b)b^n D_{lm}(b) \label{conv2}.
\eeqa
These coefficients can be themselves measured with an observationally driven Galactic model or with the {following estimators}
\beq
{\hat W}_{lmn}=\frac{1}N\sum_{i=1}^N b_i^n D_{lm}(b_i),
\eeq
 but we do not go into detail, given their simplicity.

 From the parity properties of $b^n$, $R(b)$ and $D_{lm}(b)$, we can easily derive the  key relations 
\beq
g_{lmn}=0 ~~{\rm for } ~m+l+n\equiv 1 ~({\rm mod}~2) \label{m2}.
\eeq 
For  fourfolded PDF (\ref{mul}), due to the selection rule $m\in K$, we must set $l+n=0$ (mod 2) to obtain $g_{lmn}\ne 0$. This is the reason why  we artificially added the index $n$.  We can thereby  adjust  parity relation (\ref{m2}) and measure $a_{l0}$ for an odd $l$.  As commented before, for the dipole $a_{10}$, we need to statistically amplify the small factor $D_{10}=-\sin b$, which is antisymmetric at  $b=0$ (see Fig. 2(e)). 
Indeed, taking $n=1$, we actually obtain the finite value 
\beq g_{101}=-e^{\sigma^2/2}\sigma^2.
\eeq
  For reference, we present  the leading order terms (with the expansion parameter $\sigma$) for some of nonvanishing coefficients
\beqa
g_{000}&=&1,~g_{101}=-\sigma^2,~g_{200}=-\frac12,~g_{301}=\frac32\sigma^2, \label{ex1}\\
g_{400}&=&\frac38,~
g_{440}=\frac18{\sqrt{\frac{35}{2}}}\label{ex2}
\eeqa
with the relations
$
g_{l,-mn}=(-1)^mg_{lmn}  \label{-m}$ originating from Eq. (\ref{aa10}). Note that, in \cite{Seto:2024odc}, we applied the distant observer approximation (equivalent to setting  $R(b)=\delta(b)$ and thus $\sigma=0$),  unlike the present study targeting the odd $l$ patterns.

Next,  for the estimator ${\hat X}_{lmn}$, we evaluate the shot noise $\Delta {\hat X}_{lmn}$ caused by the finiteness of the sample size $N$.   Its  variance is expressed as   
$\lla\Delta{\hat X}_{lmn} \Delta{\hat X}_{lmn}^* \rra_{I\psi b}$  and  given as 
\beq
 \frac{\lla b^n Y_{lm}^*(I,\psi)  b^n Y_{lm}(I,\psi) \rra_{I\psi b}}N\simeq \frac{\sigma^{2n}(2n-1)!!}{4\pi N}\label{shot}
\eeq 
with
$(-1)!!=1!!=1$. Here, assuming the weak signal case $|a_{l0}|\ll 1$ ($l\ge 1$), we only kept the monopole term in Eq. (\ref{pdf13}).  We  then obtain a scaling relation  for the 1-sigma detection limit  of the dipole signal, 
\beq
{\Delta a_{10}}\simeq \frac{ {\Delta {\hat X}_{101}}}{g_{101}}
\simeq 0.1   \lmk\frac{\sigma}{0.06}  \rmk^{-1}  \lmk\frac{N}{2000}  \rmk^{-1/2}.\label{sca}
\eeq 
For the quadrupole pattern $a_{20}$, we can make a similar calculation for the associated estimator ${\hat X}_{200}$  and obtain a better detection limit 
\beq
\Delta a_{20}\simeq 0.01\lmk  \frac{N}{2000}\rmk^{-1/2},\label{sca2}
\eeq
  due to  the preferable property $D_{20}(b)=O(1)$, in contrast to  $D_{10}(b)=O(b)$ (see Eq. (\ref{pdf13})).

\section{Measurement errors}
In this section, we estimate the fluctuation $\delta {\hat X}_{101}$ of the observable quantity $ {\hat X}_{101}$ due to the  estimation  errors of the involved parameters  $\lnk b_i,I_i\rnk$. Taking a linear expansion for ${\hat X}_{101}$ and assuming independence of the errors, we obtain 
\beq
\delta {\hat X}_{101}\sim \frac{|Y_{10}(\theta_i)|\delta b_i + |b_i \p_IY_{10}(\theta_i)|\delta I_i}{\sqrt N}.
\eeq
Here $\delta b_i$ and $\delta I_i$  represent the typical magnitudes of the estimation errors. Replacing the coefficients $|b_i|$, $|Y_{10}|$ and $|\p_IY_{10}|$ with their root mean square values  as $|b_i|\sim \sigma$, 
$|Y_{10}|\sim 1/\sqrt{4\pi}$ and $|\p_I Y_{10}|\sim 1/\sqrt{2\pi}$, we have 
\beq
\delta {\hat X}_{101}\sim \frac1{\sqrt{4\pi  N}}(\delta b_i+\sqrt{2} \sigma \delta I_i).\label{dx1}
\eeq

Next, requiring  that the two terms in Eq. (\ref{dx1}) are both smaller than the shot noise $
\Delta{\hat X}_{101}\simeq \sigma/{\sqrt{4\pi N} }$  given  in Eq. (\ref{shot}),
we obtain 
\beqa
\delta I_i&\lsim& 1/\sqrt2\label{in1},\\
\delta b_i&\lsim& \sigma\sim 0.06. \label{in2}
\eeqa
According to a recent numerical study on Galactic  CWDBs \cite{Zhang:2021htc}, we can expect $\delta I_i\lsim 0.2$, satisfying condition (\ref{in1}). As indicated by the Fisher matrix analysis \cite{Cutler:1994ys}, the inclination errors could be large for nearly face-on binaries.    In actual data analysis,   it might be meaningful to perform additional processing for these tail components of the inclination  estimation errors.

Meanwhile, inequality (\ref{in2}) simply indicates that we need to resolve the Galactic latitudes $b$ of  bulge  CWDBs within  the small angular width $\sigma$. In fact, this requirement will be more demanding, as we discuss below. 

To begin with,  we should recall that 
LISA revolves around the Sun on the ecliptic plane.  For a binary at the polar ecliptic coordinate $(\theta_E,\phi_E)$,  the   Doppler phase modulation induced by the revolution is  modeled as 
\beq
2\pi f t_0 \sin \theta_E\cos(\phi(t)-\phi_E).\label{dop}
\eeq 
In this expression, $t_0=500$\,s is the photon propagation time for the distance of 1a.u., and  the time dependent function $\phi_E(t)$ represents the angular position of LISA on the ecliptic plane around the Sun.

In our target frequency regime $f\gsim 4$mHz,  the Doppler modulation (\ref{dop}) usually works as 
 the primary signature for the estimation of the sky position $(\theta_E,\phi_E)$ of a binary  \cite{Cutler:1997ta}, rather than the amplitude modulation due to the time variation of the detector plane. However, for a binary around the ecliptic plane, as understood from the symmetry of expression (\ref{dop}) at $\theta_E=\pi/2$,  the error ellipse in the sky is elongated in the direction normal to the ecliptic plane \cite{Takahashi:2002ky,LISA:2022yao}.   Indeed, using the simplified Fisher matrix analysis for the gravitational wave phase of a binary \cite{Takahashi:2002ky}, the estimation error for  its angle $\theta_E$  is roughly evaluated by 
\beq
\delta \theta_{Ei}\sim \lmk \sqrt2\pi f_i t_{\rm 0} \rho_i |\cos\theta_{Ei}| \rmk^{-1}
\eeq
with the signal to noise ratio $\rho_i$ of the binary.
Unfortunately for a bulge CWDB,  the direction of the  Galactic center is only $5.6^\circ$ below  the ecliptic plane.
Considering the intersection angle $\sim 60^\circ$ between the Galactic and ecliptic planes, the estimation error for the Galactic latitude $b$ of  the binary is roughly given by
\beqa
\delta b_{i}&\sim& \frac12 \delta\theta_{Ei}\\
&\sim& 0.06\lmk \frac{\rho_i}{10} \rmk^{-1} \lmk \frac{f_i}{\rm 4mHz} \rmk^{-1} \lmk \frac{\cos \theta_{Ei}}{0.1} \rmk^{-1} \label{in33}.
\eeqa
While the amplitude modulation could reduce the estimation error $\delta b_i$ to some extent,
given condition (\ref{in2}), we might need to  increase the product $f_i \rho_i$ by taking a larger frequency threshold at the sample selection for the estimator ${\hat X}_{101}$. We might also need to  reduce the statistical weights for binaries very close to the ecliptic plane $\theta_E=\pi/2$. These treatments could reduce the effective sample size $N$.  

Note that the estimator ${\hat X}_{200}$ for the quadrupole pattern $a_{20}$ does not directly depend on  the Galactic latitude $b$. From the comparison between the effects of  the shot noise and the parameter estimation errors, we  obtain  a requirement similar to inequality (\ref{in1}).

\section{extra-Galactic studies with  BBO/DECIGO}

Next, as an application of our approach, we discuss the extra-Galactic binary analysis. BBO \cite{Crowder:2005nr,Harry:2006fi} and DECIGO \cite{Seto:2001qf,Kawamura:2011zz} are proposed  space missions after LISA, primarily targeting  primordial  GW backgrounds from the early universe around 0.1-10Hz. We need to individually detect and subtract foreground inspiral signals made by  black hole binaries  (BHBs) and  neutron star binaries  (NSBs) at cosmological distances.  Using widely ($\sim$1a.u.) separated multiple units, these missions are designed to realize good sky localization for the binaries (see also \cite{Yagi:2011wg}). Indeed,   the expected number of host galaxies in the error cube of BBO is $\sim 0.01$ for a BHB at $z\sim 1$ \cite{Cutler:2009qv}. The host galaxy (and its redshift) determination  is  critical for high-precision cosmology and  could also allow us to estimate its rotation axis $\vec q$. Interestingly, two plausible solutions (geometrically corresponding to the polarization angles $\psi$ and $\psi+\pi$) appear again,  but now for the  rotational vectors $\vec q$ of distant spiral galaxies.  This issue is well known in relation to the census of the trailing and leading spiral arm patterns (see e.g.  \cite{gd,dev,iye}). The empirical laws (e.g. almost always trailing  \cite{gd,iye}) will help us to efficiently select the right solution for $\vec q$. 

With the recent estimation for the BHB merger rate $\rm \sim 24 Gpc^{-3}yr^{-1}$ \cite{LIGOScientific:2020kqk}, the detection rate  for BHBs at $z\lsim 1$ will be $\sim 4000{\rm yr^{-1}}$.  
Unlike small offset angles $|b|\ll 1$ for Galactic CWDBs, the  host galaxies of the BHBs will have have  random orientations $\vec q$ relative to their directions $\vec n$ (corresponding to typical offset angles of $b=O(1)$).
Thus, the detection limits for both the dipole and quadrupole patterns ($l=1$ and 2) will be $\Delta a_{l0}\sim 0.01(N/2000)^{-1/2}$. By increasing the observational period and thus the sample size $N$, we might detect the alignment signatures, which  will be quite useful to argue the formation scenarios of BHBs.

A NSB would have an  error cube  one order of magnitude larger than that of a BHB \cite{Cutler:2009qv}, but we might observe EM signals which can be  predicted ahead of the NSB  merger.  It should  also be noted that the kick velocities at the formations of neutron stars could perturb the orbital orientations $\vec j$ from the initial ones.

\section{Summary}

In the previous paper \cite{Seto:2024odc},
we discussed the orientational analysis of bulge CWDBs with the planned GW detector LISA, ignoring the bulge thickness (equivalent to $\sigma=0$).
This simple approximation symmetrically cancelled the signature of the odd-$l$ patterns. 
 In this paper,   paying attention to the dipole pattern measurement,  we developed a new formulation to deal with the projected  bulge structure in the sky.  We also introduced the refined estimators ${\hat X}_{lmn}$ for separately measuring the observable harmonic patterns. Here  the index $n$ is introduced  for leveraging the parity properties of the system.
We then evaluated the shot noises  and obtained  simple scaling relations (\ref{sca}) and  (\ref{sca2}), which will help us to  assess the prospects for measuring the low $l$ patterns. 
{Since the direction of the Galactic center is close to the ecliptic plane,  it might be meaningful to develop a refined analysis scheme for mitigating the relatively poor sky localization of the bulge CWDBs. }
 finally, we presented a brief discussion on the extra-Galactic studies with BBO and DECIGO. 

\acknowledgements
{The author would like to thank the reviewer for useful comments.}




\end{document}